\documentclass[preprint,prc,showpacs,preprintnumbers,amsmath,amssymb,floatfix]{revtex4}
\usepackage{amsmath}

\usepackage{graphicx}
\usepackage{dcolumn}
\usepackage{bm}
\usepackage{ulem} 
\usepackage[usenames]{color}
\usepackage{epstopdf}
\usepackage{epsfig}
\usepackage{float}
\usepackage{subfigure}

\newcommand{\nc}{\newcommand}       
\nc{\vc}[1] {\mbox{\boldmath $#1$}} 
\nc{\del}       {\partial}              
\nc{\bra}       {\langle}               
\nc{\ket}       {\rangle}               
\nc{\bras}[1]   {\langle #1|}           
\nc{\kets}[1]   {|#1\rangle}            
\nc{\mapleft}[1]{           
 \smash{\mathop{\,          %
  \hbox to 1.5cm{\rightarrowfill}\, }\limits_{#1}}}
\nc{\beq}     {\begin{eqnarray}} \nc{\eeq}    {\end{eqnarray}}
\nc{\nn}      {\\\nonumber} \nc{\vs}      {\vspace{-0.275cm}}
\nc{\fra}    {\frac{1}{2}}
\nc{\mb}        {\mathbf}

\begin{document}

\preprint{}

\title{The phase transition in hot $\Lambda$ hypernuclei within relativistic Thomas-Fermi approximation}

\author{Jinniu Hu}
\email{hujinniu@nankai.edu.cn}
\affiliation{School of Physics, Nankai University, Tianjin 300071,  China}
\author{Zhaowen Zhang}
\affiliation{Department of Physics and Astronomy and Shanghai Key Laboratory for Particle Physics and Cosmology, Shanghai Jiao Tong University, Shanghai 200240, China}
\author{Shishao Bao}
\affiliation{School of Physics, Nankai University, Tianjin 300071,  China}
\author{Hong Shen}
\affiliation{School of Physics, Nankai University, Tianjin 300071,  China}

\date{\today}
\begin{abstract}
A self-consistent description for hot $\Lambda$ hypernuclei in hypothetical big boxes is developed within the relativistic Thomas-Fermi approximation in order to investigate directly the liquid-gas phase coexistence in strangeness finite nuclear systems. We use the relativistic mean-field model for nuclear interactions.  The temperature dependence of $\Lambda$ hyperon density, $\Lambda$ hyperon radius, excitation energies, specific heat, and the binding energies of $\Lambda$ hypernuclei from $^{16}_{\Lambda}$O to $^{208}_{\Lambda}$Pb in phase transition region are calculated by using the subtraction procedure in order to separate the hypernucleus from the surrounding baryon gas. The $\Lambda$ central density is very sensitive to the temperature. The radii of $\Lambda$ hyperon at high temperature become very large. In the relativistic Thomas-Fermi approximation with the subtraction procedure, the properties of hypernuclei are independent of the size of the box in which the calculation is performed. The level density parameters of hypernuclei in the present work are confirmed to be almost constant at low temperature. It is also found that the single-$\Lambda$ binding energies of $\Lambda$ hypernuclei are largely reduced with increasing temperature.
\end{abstract}



\maketitle

\section{Introduction}

Theoretical studies of hypernuclei are continuously boosted by new and upgraded experimental facilities~\cite{LL1,LL2,LL3,LL4,LL5,Gal10,Bot12,Tam12}. It is generally believed that from them one could derive various features of the underlying hyperon interactions~\cite{Bod87,Usm04,Usm06,Hiy10,Hiy10s,Hiy12,Gal11,Li13}. They are also related to the dense stellar matter studies~\cite{Li07,Hu13}, as an alternative way of obtaining the matter apart from astrophysical observations and/or quite developed many-body schemes for infinite strongly interacting systems, for example, the widely used microscopic Brueckner-Hartree-Fock (BHF) theory~\cite{Bur11}.

Lattice QCD calculation should be an ideal tool for investigating hypernucleus structure since it retains all the fundamental characters of QCD theory. Indeed, the first calculation of hypernuclei with baryon number $A > 2$ has been performed recently, for $^4_{\Lambda}$He and $^4_{\Lambda\Lambda}$He~\cite{lqcd}. However, a detailed and precise structure description is still beyond its reach. Few-body calculations in cluster or shell-model approach are awaited for not-so-light hypernuclei ($A > 10$). Significant progress in the auxiliary field diffusion Monte Carlo method~\cite{Lon13} has been achieved in the calculation of closed shell $\Lambda$ hypernuclei from $A = 5$ to $91$. For a more feasible way of the systematic study of both light and heavy hypernuclei, effective models are generally employed. Among them, many models are for single-$\Lambda$ hypernuclei, for example, the quark mean-field model~\cite{shen02}, the relativistic mean-field (RMF) approach~\cite{shen06,Xu12}, the Skyrme-Hartree-Fock model~\cite{Li13,Sch13,Gul12}, the quark-meson coupling model~\cite{Gui08}, a relativistic point-coupling model~\cite{Tan12}, the quark mass density-dependent model~\cite{Wu13}, and the density-dependent RMF theory from relativistic BHF theory \cite{Hu14}.

The experiments, $(\pi,K)$, $(e,e'K)$ and $(\gamma, K)$, are the most popular reactions used to produce hypernuclei ~\cite{hashimoto06}. Recently, the heavy ion collision is suggested as one way to generate hypernuclei ~\cite{gaitanos09,botvina11}, such as the high energy Au+Au collision~\cite{steinheimer12}, which can be considered as a liquid-gas phase transition in hypermatter. The lifetime of hypernuclei in such reactions are usually very short and the production of hypernuclei should be strongly dependent on the temperature. Therefore, it is very interesting to investigate the properties of hot hypernuclei in the liquid-gas coexistence region. The matter generated from the collision of relativistic heavy ions has some probabilities to break up as the nuclear-fragment and hyper-fragment production, which can be described by the statistical multifragmentation model~\cite{bondorf95}. This model was also extended to the study of hypernuclei produced in heavy ion collisions~\cite{botvina07,gupta09}. 

Accordingly, we want to investigate the hot hypernuclei from the aspect of the liquid-gas phase coexistence in this work. Since  the hot hypernucleus formed in nucleus-nucleus collisions is thermodynamically unstable against the emission of baryons, an external pressure has to be exerted on the hypernucleus to compensate for the tendency of baryon emission. This pressure is assumed to be exerted by a surrounding gas representing evaporated baryons, which is in equilibrium with the hot hypernucleus. 

In order to separate the nucleus from the surrounding gas, a subtraction procedure was first proposed in Hartree-Fock framework~\cite{BLV85} for normal nucleus, and then used in the Thomas-Fermi approach~\cite{TF87}. The subtraction procedure is based on the existence of two solutions to the equations of motion of nucleons. One solution corresponds to the nucleon gas alone ($G$), and the other to the nuclear liquid phase in equilibrium with the surrounding gas ($NG$). The density profile of the nucleus ($L$) is then given by subtracting the gas density from that of the liquid-plus-gas phase. Finally, the physical quantities of the isolated nucleus obtained using such subtraction procedure could be independent of the size of the box in which the calculation is performed. In the past decades, this subtraction procedure has been widely applied in the non-relativistic Thomas-Fermi approximation with Skyrme force~\cite{Sub01,Sub02,Sub07,Sub12,PLB12,Sub14}.

The relativistic Thomas-Fermi approximation with RMF Lagrangian has been developed and applied to study various subjects at the subnuclear densities, such as, droplet formation~\cite{RTF99a,RTF99b} and nuclear pasta phases~\cite{Mene08,Mene10,Gril12}. This method is considered to be self-consistent in the treatment of surface effects and nucleon distributions. The relativistic Thomas-Fermi approximation was also adopted to describe finite nuclei~\cite{RTF02,RTF01} and non-uniform nuclear matter for supernova simulations~\cite{Zhang14}. In Refs.~\cite{RTF02,RTF01}, the thermodynamic properties of finite nuclei were calculated within the relativistic Thomas-Fermi approximation, and the results obtained were found to depend on the input freeze-out volume, which was actually the size of the box for performing the calculation. Recently, we developed a relativistic Thomas-Fermi model for the description of hot nuclei by employing the subtraction procedure, and investigated the temperature dependence of the symmetry energy of finite nuclei \cite{zhang14b}. Actually, the results obtained from subtraction procedure are independent of the size of the box.

In this work, we would like to extend the relativistic Thomas-Fermi model with subtraction procedure to describe the hot $\Lambda$ hypernuclei, which are most known in experiment and theoretical calculation among various hypernuclei. For the nuclear interaction and $\Lambda N$ interaction, we adopt the RMF model, which has been successfully used to study various phenomena in nuclear physics~\cite{Sero86,Ring90,Meng06}. The thermodynamic properties of hot $\Lambda$ hypernuclei, such as excitation energies, specific heat, and level density parameters of hypernuclei will be investigated.

In Sec. II, we briefly derive the relativistic Thomas-Fermi approximation using the subtraction procedure for the description of hot $\Lambda$ hypernuclei. In Sec. III, the numerical results are shown for the properties of $\Lambda$ hypernuclei from $^{16}_{\Lambda}$O to $^{208}_{\Lambda}$Pb at finite temperature. A summary is given in Sec. IV.

\section{Relativistic Thomas-Fermi approximation for hot $\Lambda$ hypernuclei}
In the RMF model, the baryons (nucleons and hyperons) interact through the exchange of various mesons. The mesons considered are the isoscalar scalar and vector mesons ($\sigma$ and $\omega$) and isovector vector meson ($\rho$). The baryon Lagrangian density reads,
\beq
\label{eq:LRMF}
\mathcal{L}_{\rm{RMF}} & = & \sum_{i=p,n}\bar{\psi}_i
\left[i\gamma_{\mu}\partial^{\mu}-M_i-g_{\sigma N}\sigma
      -g_{\omega N}\gamma_{\mu}\omega^{\mu}-\frac{g_{\rho N}}{2}\gamma_{\mu}\tau_a\rho^{a\mu}
      -e\gamma_\mu\frac{(1-\tau_3)}{2}A^\mu \right]\psi_i \nn
&&+\bar{\psi}_\Lambda ( i\gamma_{\mu}\partial^{\mu}- M_\Lambda- g_{\sigma \Lambda}\sigma- g_{\omega
   \Lambda}\gamma_{\mu}\omega^{\mu}){\psi}_\Lambda \nn
&& +\frac{1}{2}\partial_{\mu}\sigma\partial^{\mu}\sigma
   -\frac{1}{2}m^2_{\sigma}\sigma^2-\frac{1}{3}g_{2}\sigma^{3}
   -\frac{1}{4}g_{3}\sigma^{4} \notag \nn
&& -\frac{1}{4}W_{\mu\nu}W^{\mu\nu} +\frac{1}{2}m^2_{\omega}\omega_{\mu}\omega^{\mu}
   +\frac{1}{4}c_{3}\left(\omega_{\mu}\omega^{\mu}\right)^2  \notag \nn
&& -\frac{1}{4}R^a_{\mu\nu}R^{a\mu\nu} +\frac{1}{2}m^2_{\rho}\rho^a_{\mu}\rho^{a\mu},
\eeq
where $W^{\mu\nu}$ and $R^{a\mu\nu}$ are the antisymmetric field tensors for $\omega^{\mu}$ and $\rho^{a\mu}$, respectively. $g_{\sigma N},~g_{\omega N}$ and $g_{\rho N}$ are the coupling constants between $\sigma, \omega, \rho$ and nucleon, respectively, while $g_{\sigma \Lambda}$ and $g_{\omega \Lambda}$ are the coupling constants between $\sigma,~\omega$ and $\Lambda$ hyperon. Here, the tensor coupling between $\omega$ and $\Lambda$ hyperon is not taken into account, which just generates the large spin-orbit splitting of $\Lambda$ hyperon. However, in the Thomas-Fermi approximation, the single particle level at different spin states cannot be obtained. Furthermore, such tensor coupling does not change the total energy of $\Lambda$ hypernuclei very much. Therefore, we ignore this tensor coupling term in present work. The electromagnetic coupling constant is $e=\sqrt{4\pi/137}$. In the RMF approach, meson fields are treated as classical fields and the field operators are replaced by their expectation values. For a static system, the nonvanishing expectation values are $\sigma =\left\langle \sigma \right\rangle$, $\omega =\left\langle \omega^{0}\right\rangle$, and $\rho =\left\langle \rho^{30} \right\rangle$, where $0$ represents the time component in Dirac space and $3$ represents the third component in isospin space for $\rho$ meson.

Using the relativistic Thomas-Fermi approximation with the subtraction procedure \cite{BLV85,TF87}, we study a hot $\Lambda$ hypernucleus based on the thermodynamic potential of the isolated hypernucleus, which is defined by
\beq\label{eq:ON}
\Omega=\Omega^{NG}-\Omega^{G}+E_{C},
\eeq
where $\Omega^{NG}$ and $\Omega^{G}$ are the baryon thermodynamic potentials in the liquid phase with the surrounding gas ($NG$) and the gas phase alone ($G$), respectively. We employ the RMF Lagrangian to obtain the thermodynamic potential $\Omega^{a}$ ($a=NG$ or $G$), which can be given as
\beq
\label{eq:Ormf}
\Omega^a=E^a-TS^a-\sum_{i=p,n,\Lambda}\mu_{i}N^a_{i}.
\eeq
Here, the energy $E^a$, entropy $S^a$, and particle number $N^a_i$ in the phase $a$ are obtained by
\beq\label{eq:Ea}
E^a &=&\int \varepsilon^a (r)d^3 r,\nn
S^a &=&\int s^a (r) d^3 r,\nn
N^a_{i} &=&\int n^a_{i}(r) d^3 r,
\eeq
where $\varepsilon^a(r)$,  $s^a(r)$, and $n^a_{i}(r)$ are the local energy density, entropy density, and particle number density defined in the RMF model. The local energy density derived from the Lagrangian density~(\ref{eq:LRMF}) without Coulomb force is written as
\beq\label{eq:ermf}
{\varepsilon}(r) &=&\displaystyle{\sum_{i=p,n,\Lambda}
\frac{1}{\pi^{2}}\int_{0}^{\infty}dk\,k^{2}\,\sqrt{k^{2}+{M_i^{\ast }}^{2}}
\left( f_{i+}^{k}+f_{i-}^{k}\right) } \nn
&&+\frac{1}{2}(\nabla \sigma )^{2}+\frac{1}{2}m_{\sigma }^{2}\sigma ^{2}+%
\frac{1}{3}g_{2}\sigma ^{3}+\frac{1}{4}g_{3}\sigma ^{4}  \nn
&&-\frac{1}{2}(\nabla \omega )^{2}-\frac{1}{2}m_{\omega }^{2}\omega ^{2}-%
\frac{1}{4}c_{3}\omega ^{4}+g_{\omega N }\omega \left( n_{p}+n_{n}\right)+g_{\omega\Lambda }\omega n_{\Lambda}\nn
&&-\frac{1}{2}(\nabla \rho )^{2}-\frac{1}{2}m_{\rho N }^{2}\rho ^{2}+\frac{%
g_{\rho }}{2}\rho \left( n_{p}-n_{n}\right),
\eeq
where $M_i^{\ast}=M_i+g_{\sigma i}\sigma$ is the effective baryon mass, and $n_{i}$ is the number density of species $i$ ($i=p,n$ or $\Lambda$).
The entropy density is given by
\beq
\label{eq:srmf}
s(r)=\displaystyle{\sum_{i=p,n,\Lambda}\frac{1}{\pi^{2}} \int_{0}^{\infty}dk\,k^{2}}
& \left[ -f_{i+}^{k}\ln f_{i+}^{k}-\left( 1-f_{i+}^{k}\right) \ln \left(
1-f_{i+}^{k}\right) \right.  \nn
& \left. -f_{i-}^{k}\ln f_{i-}^{k}-\left( 1-f_{i-}^{k}\right) \ln \left( 1-f_{i-}^{k}\right) \right].
\eeq
Here $f_{i+}^{k}$ and $f_{i-}^{k}$ are the occupation probabilities of the particle and antiparticle at momentum $k$, respectively. Their detailed form will be determined by variational principle self-consistently later.
The number density of proton ($i=p$), neutron ($i=n$) or $\Lambda$ hyperon ($i=\Lambda$) at position $r$ is given by
\beq
\label{eq:nirmf}
 n_{i}(r)=\frac{1}{\pi^2}
       \int_0^{\infty} dk\,k^2\,\left(f_{i+}^{k}-f_{i-}^{k}\right).
\eeq
The Coulomb energy is calculated from the subtracted proton density as
\beq
\label{eq:EC}
E_{C}=\int \left[ e \left( n_{p}^{NG}-n_{p}^{G}\right) A_0
                  -\frac{1}{2}(\nabla A_0 )^{2}\right] d^{3}r,
\eeq
where $A_0$ is the electrostatic potential.

The equilibrium state of the isolated hypernucleus can be obtained by minimization of the thermodynamic potential $\Omega$ defined in Eq.~(\ref{eq:ON}). The meson mean fields in the $NG$ phase satisfy the variational equation
\beq
\label{eq:MNG}
\frac{\delta \Omega }{\delta \phi^{NG}}=0,
\hspace{1cm} \phi^{NG} =\sigma^{NG},\,\omega^{NG},\,\rho^{NG},
\eeq
which leads to the following equations of motion for meson mean fields in the $NG$ phase,
\beq\label{eq:png}
&&-\nabla^{2}\sigma^{NG}+m_{\sigma}^{2}\sigma^{NG}
+g_{2}\left(\sigma^{NG}\right)^{2}+g_{3}\left(\sigma^{NG}\right)^{3}
=-g_{\sigma N}\left(n_{s,p}^{NG}+n_{s,n}^{NG}\right)-g_{\sigma\Lambda}n_{s,\Lambda}^{NG}, \nn
&&-\nabla^{2}\omega^{NG}+m_{\omega}^{2}\omega^{NG}
+c_3\left(\omega^{NG}\right)^{3}
=g_{\omega N}\left(n_{p}^{NG}+n_{n}^{NG}\right)+g_{\omega\Lambda}n_{\Lambda}^{NG} , \nn
&&-\nabla^{2}\rho^{NG}+m_{\rho}^{2}\rho^{NG}
=\frac{g_{\rho N}}{2}\left( n_{p}^{NG}-n_{n}^{NG}\right) .
\eeq
The occupation probability $f_{i+}^{k,NG}$ ($f_{i-}^{k,NG}$) of species $i$ ($i=p,n$ or $\Lambda$) can be derived from the variational equation,
\beq
\frac{\delta \Omega }{\delta f_{i\pm}^{k,NG}}=0,
\eeq
which results in the Fermi-Dirac distribution of particle and antiparticle for proton or neutron as,
{\footnotesize
\beq
f_{i\pm}^{k,NG} &=&\left\{1+\exp \left[ \left( \sqrt{k^{2}+
 \left( M_i^{\ast,NG}\right)^{2}}+g_{\omega N }\omega^{NG}
 +\frac{g_{\rho N }}{2}\tau_{3}\rho^{NG}+e\frac{\tau_{3}+1}{2}A_0
 \mp \mu_{i}\right) /T\right] \right\}^{-1},
\eeq}
and the one for $\Lambda$ hyperon
\beq
f_{\Lambda\pm}^{k,NG} &=&\left\{1+\exp \left[ \left( \sqrt{k^{2}+
 \left( M_\Lambda^{\ast,NG}\right)^{2}}+g_{\omega \Lambda }\omega^{NG}
 \mp \mu_{\Lambda}\right) /T\right] \right\}^{-1}.
\eeq
Similarly, we obtain the equations of motion for meson mean fields in the $G$ phase,
\beq\label{eq:pg}
&&-\nabla^{2}\sigma^{G}+m_{\sigma}^{2}\sigma^{G}
+g_{2}\left(\sigma^{G}\right) ^{2}+g_{3}\left(\sigma^{G}\right)^{3}
=-g_{\sigma N }\left(n_{s,p}^{G}+n_{s,n}^{G}\right)-g_{\sigma\Lambda}n_{s,\Lambda}^{G}, \nn
&&-\nabla^{2}\omega^{G}+m_{\omega}^{2}\omega^{G}
+c_3\left(\omega^{G}\right)^{3}
=g_{\omega N}\left(n_{p}^{G}+n_{n}^{G}\right)+g_{\omega \Lambda}n_{\Lambda}^{G}, \nn
&&-\nabla^{2}\rho^{G}+m_{\rho}^{2}\rho^{G}
=\frac{g_{\rho N}}{2}\left( n_{p}^{G}-n_{n}^{G}\right),
\eeq
and the occupation probability in the $G$ phase for proton or neutron,
{\small
\beq
f_{i\pm}^{k,G} &=&\left\{1+\exp \left[ \left( \sqrt{k^{2}+
 \left( M_i^{\ast,G}\right)^{2}}+g_{\omega N}\omega^{G}
 +\frac{g_{\rho N}}{2}\tau_{3}\rho^{G}+e\frac{\tau_{3}+1}{2}A_0
 \mp \mu_{i}\right) /T\right] \right\}^{-1},
\eeq
}
and the one for $\Lambda$ hyperon,
{\small
\beq
f_{\Lambda\pm}^{k,G} &=&\left\{1+\exp \left[ \left( \sqrt{k^{2}+
 \left( M_\Lambda^{\ast,G}\right)^{2}}+g_{\omega \Lambda}\omega^{G}
 \mp \mu_{\Lambda}\right) /T\right] \right\}^{-1}.
\eeq
}
In the equations for meson mean fields, $n_{s,i}^{a}$ and $n_{i}^{a}$ denote respectively the scalar and number densities of species $i$ ($i=p,n$ or $\Lambda$) in the $a$ ($a=NG$ or $G$) phase \cite{zhang14b}. By minimizing $\Omega$ with respect to the electrostatic potential $A_0$,
we obtain the Poisson equation for  $A_0$ as
\beq
\label{eq:EQA}
-\nabla^{2}A_0=e\left( n_{p}^{NG}-n_{p}^{G}\right) .
\eeq
The inclusion of the Coulomb energy in $\Omega$ leads to a coupling between the two sets of equations for the $NG$ and $G$ phases. Therefore, the coupled equations (\ref{eq:png}), (\ref{eq:pg}), and~(\ref{eq:EQA}) should be solved simultaneously at given temperature $T$ and chemical potentials $\mu_p,~\mu_n$ and $\mu_\Lambda$.

For a hypernucleus with $N_p$ protons, $N_n$ neutrons and $N_\Lambda$ hyperons at temperature $T$, the proton, neutron and $\Lambda$ hyperon chemical potentials $\mu_p$, $\mu_n$ and $\mu_\Lambda$ can be determined from given $N_p$, $N_n$ and $N_\Lambda$. Once the chemical potentials are known, the occupation probabilities and density distributions can be obtained easily. In practice, we solve self-consistently the coupled equations (\ref{eq:png}), (\ref{eq:pg}), and~(\ref{eq:EQA}) under the constraints of given $N_p,~N_n$ and $N_\Lambda$. After getting the solutions for the $NG$ and $G$ phases, we can extract the properties of the hot hypernucleus based on the subtraction procedure. The proton, neutron, and $\Lambda$ hyperon numbers, $N_p, ~N_n$ and $N_\Lambda$, are given by
\beq
N_{i} = N^{NG}_i - N^{G}_i=\int n_i(r) d^3 r,
\hspace{1cm} i=p,\, n,\, \Lambda,
\label{eq:Ni}
\eeq
where $n_{i}(r)=n^{NG}_i(r)-n^{G}_i(r)$ is the local density of the isolated hypernucleus, which decreases to zero at large distances. Therefore, physical quantities of the isolated hypernucleus could be independent of the size of the box in which the calculation is done. The total energy including Coulomb contributions for the hot hypernucleus is given by
\beq
\label{eq:EN}
E=E^{NG}-E^{G}+E_{C},
\eeq
where $E^{NG}$ and $E^{G}$ are the baryon energies without Coulomb interaction in the $NG$ and $G$ phases, which are calculated from Eq.~(\ref{eq:Ea}). The Coulomb energy $E_{C}$ is given by Eq.~(\ref{eq:EC}). The entropy and other extensive quantities of the isolated hypernucleus can be calculated by subtracting the contribution of the $G$ phase from the one of the $NG$ phase.

The excitation energy of hot hypernuclei is a very important thermodynamic quantity. For a hypernucleus at temperature $T$, its excitation energy is defined as
\beq\label{eq:Estar}
E^{\ast}(T)=E(T)-E(T=0).
\eeq
The center-of-mass correction of $\Lambda$ hypernucleus is taken into account by a conventional phenomenological way \cite{Xu12},
\beq
E_{\textrm{c.m.}}=-\frac{3}{4}\times41 \left(N_n+N_p+N_\Lambda\right)^{-1/3} \text{MeV}.
\eeq

\section{Results and discussions}
The properties of hot $\Lambda$ hypernuclei are investigated within the relativistic Thomas-Fermi approximation using the subtraction procedure in this section. For the nuclear interaction, we adopt the RMF model with TM1 parametrization \cite{TM1}, which was determined by the ground-state properties of finite nuclei and properties of nuclear matter from relativistic BHF theory. It was successfully applied to calculate the equation of state for supernova simulations and characters of neutron stars \cite{Shen11,Shen02}. As for the meson-$\Lambda$ hyperon couplings, it is well known that the properties of $\Lambda$ hypernuclei are very sensitive to the ratios of the meson-$\Lambda$ hyperon couplings to the meson-nucleon couplings $R_\sigma=g_{\sigma\Lambda}/g_{\sigma N}$ and $R_\omega=g_{\omega\Lambda}/g_{\omega N}$.
We take the relative $\omega$ coupling as $R_\omega=2/3$ from the naive quark counting and the relative $\sigma$ coupling as $R_\sigma=0.621$ given in Ref. \cite{shen06}. With this choice, the experimental $\Lambda$ binding energies of single-$\Lambda$ hypernuclei can be reproduced very well in the RMF model \cite{shen06}.

\begin{figure}[!hbt]
\includegraphics[bb=0 0 240 270, width=0.5\textwidth]{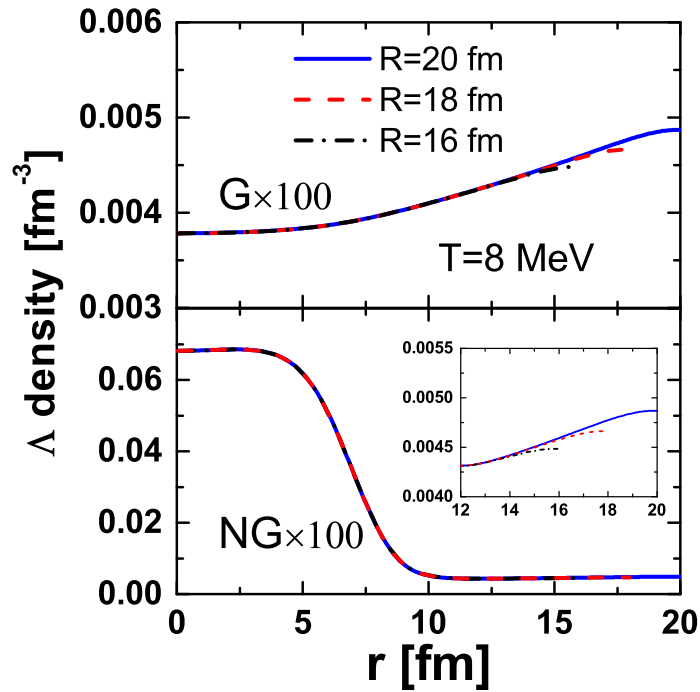}
\caption{(Color online) The density distributions of $\Lambda$ hyperon for $^{208}_{\Lambda}$Pb at $T=8$ MeV obtained with different box sizes $R=16,~18$, and $20$ fm. The density distributions from the gas phase ($G$) and the liquid-plus-gas phase ($NG$) are shown in the top and bottom panels, respectively.}\label{box}
\end{figure}

The coupled equations (\ref{eq:png}), (\ref{eq:pg}) and (\ref{eq:EQA}) are solved self-consistently with given baryon numbers of $\Lambda$ hypernuclei, $N_n,~N_p$ and $N_\Lambda$ from Eq.~(\ref{eq:Ni}) in a spherical box with radius $R$. In this section, we take two single-$\Lambda$ hypernuclei, $^{40}_{\Lambda}$Ca and $^{208}_{\Lambda}$Pb, as numerical examples to investigate the properties of hot hypernuclei within relativistic Thomas-Fermi approximation. In the subtraction procedure, the properties of hot hypernuclei should be independent of the size of the box, when the box radius $R$ is generally taken to be sufficiently large. In Fig. \ref{box}, the density distributions of $\Lambda$ hyperon from $^{208}_{\Lambda}$Pb for $G$ and $NG$ phases at $T=8$ MeV with different box sizes, $R=16,~18$, and $20$ fm are shown in order to check if the results depend on the size of the box. At the central region of the hypernucleus, these distributions are identical, while they have different behaviors approaching the box boundary. However, the behaviors of the $G$ phase at boundary are in accordance with the one of the $NG$ phase, which will generate their subtraction, i.e. the densities of the $L$ phase, to be independent of the size of the box.

\begin{figure}[!hbt]
\includegraphics[bb=20 60 310 325, width=0.8\textwidth]{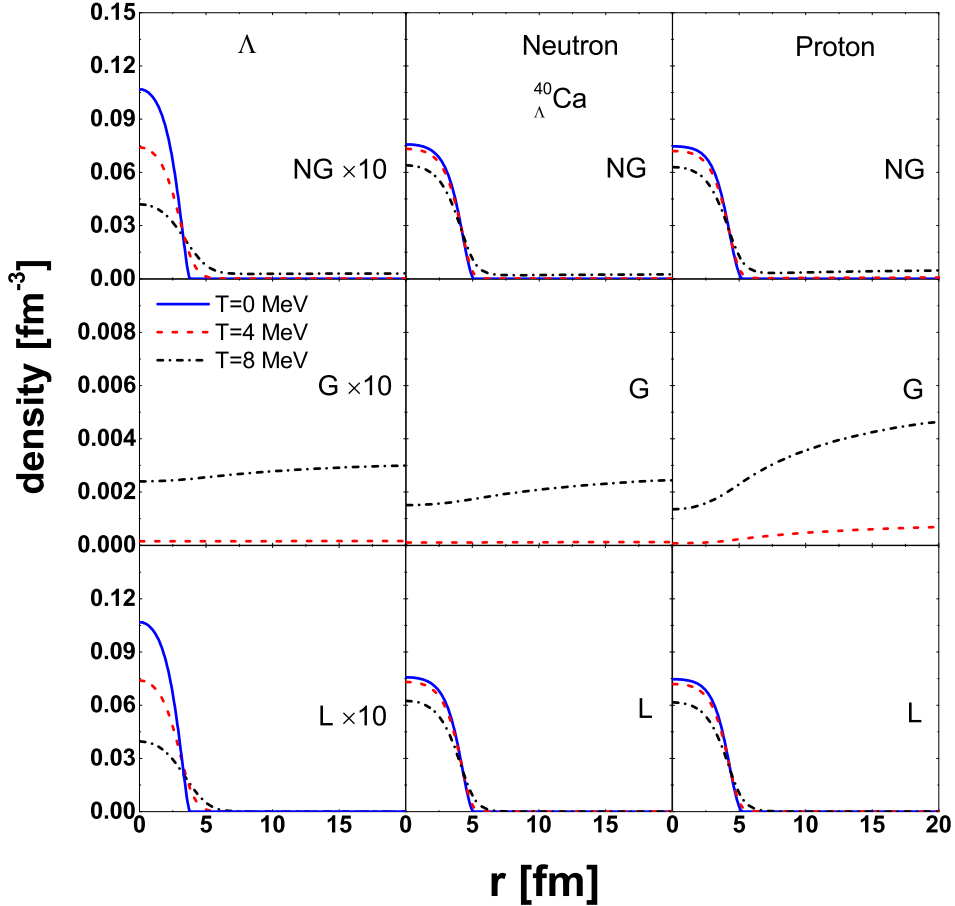}
\caption{(Color online) The density distributions of $\Lambda$ hyperon (left panels), neutron (middle panels), and proton (right panels) for $^{40}_{\Lambda}$Ca at $T=0,~4,$ and $8$ MeV obtained using the TM1 parametrization. The density distributions from liquid-plus-gas ($NG$), gas phase ($G$), and subtracted liquid phase ($L$) are shown in the top, middle and bottom panels, respectively.}\label{cad}
\end{figure}

\begin{figure}[!hbt]
\includegraphics[bb=20 60 310 325, width=0.8\textwidth]{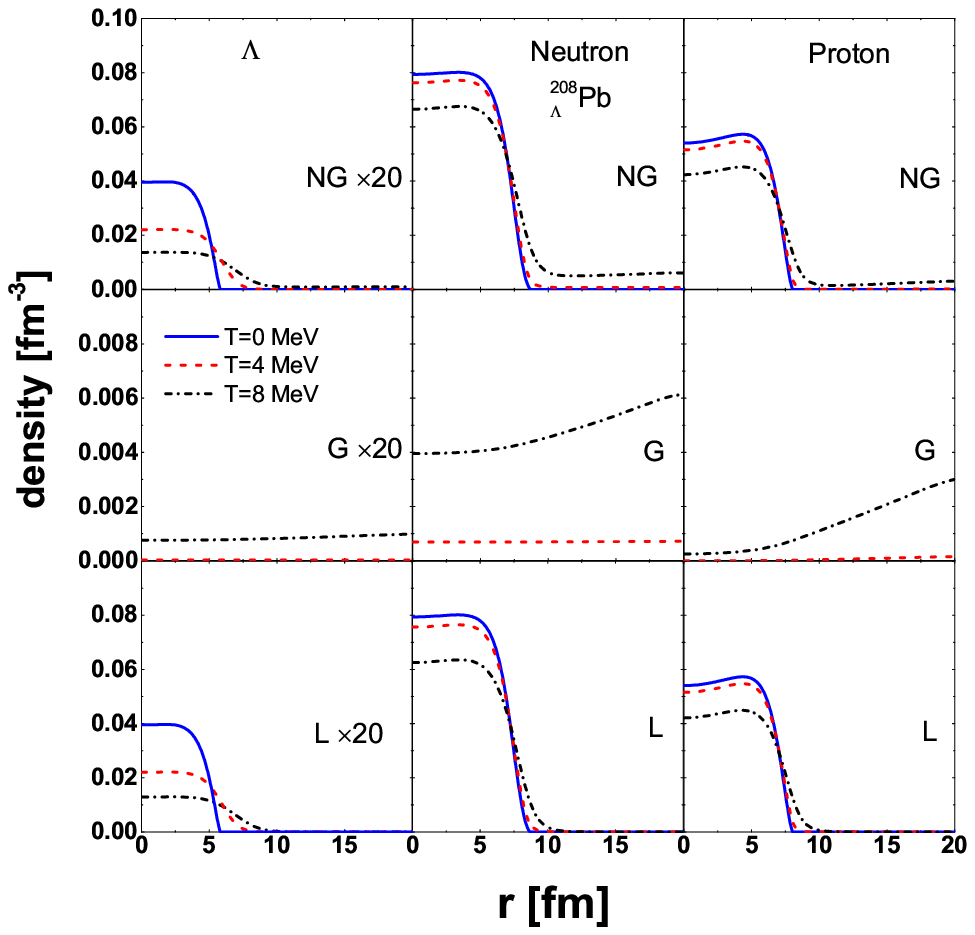}
\caption{(Color online) Same quantities as Fig.~\ref{cad}, but for $^{208}_{\Lambda}$Pb. }\label{pbd}
\end{figure}

In Figs. \ref{cad} and \ref{pbd}, the density distributions of $\Lambda$ hyperon, neutron and proton for $^{40}_{\Lambda}$Ca and $^{208}_{\Lambda}$Pb at $T=0,~4$, and $8$ MeV from left panels to right panels are presented, which are obtained with the box radius, $R=20$ fm. From top to bottom, the results of the liquid-plus-gas phase ($NG$), gas phase ($G$), and subtracted liquid phase ($L$) are displayed, respectively. The $\Lambda$ hyperon density distributions are multiplied by 10 and 20 in $^{40}_{\Lambda}$Ca and $^{208}_{\Lambda}$Pb, respectively to adopt the same scales of neutron and proton in these two figures. Firstly, we can see that subtracted $\Lambda$ densities in the isolated hypernucleus ($L$) vanish at large distances. Therefore, the physical quantities of the hypernucleus will be independent of the size of the box. The $\Lambda$ densities of the $G$ phase are found to be exactly zero at zero temperature, while these densities are finite but very small at low temperature ($T=4$ MeV). As temperature increases, the $\Lambda$ hyperon densities of the $G$ phase increase obviously. On the other hand, the $\Lambda$ densities at the center of the hypernucleus are reduced largely and the nuclear surface becomes more diffuse with increasing $T$ as shown in the top and bottom panels. The $\Lambda$ hyperon density in the center region at $T=8$ MeV is just about $30\%$ of the value at $T=0$ MeV. It is easier to be influenced by the temperature for a single-$\Lambda$ hyperon compared with a large nucleus composed of many protons and neutrons whose center densities are less sensitive to the temperature as shown in Figs. 1 and 2 in Ref.~\cite{zhang14b}. Moreover, the $\Lambda$ density at the center of $^{40}_{\Lambda}$Ca is about 5 times of the one of $^{208}_{\Lambda}$Pb. This is because the $\Lambda$ density in a single-$\Lambda$ hypernucleus is inversely proportional to the baryon number, $n_\Lambda \propto \frac{1}{A}$, if we consider the hypernucleus as a liquid drop. For the neutron and proton densities in $^{40}_{\Lambda}$Ca and $^{208}_{\Lambda}$Pb, they were almost not changed by $\Lambda$ hyperon compared with the ones of $^{39}$Ca and $^{207}$Pb without $\Lambda$ hyperon as shown in Ref.~\cite{zhang14b}.  This is because that the magnitude of single $\Lambda$ hyperon density is very small, just $5\%\sim10\%$ of nucleons. It will not change the solutions of Eq.(\ref{eq:png}) and Eq.(\ref{eq:pg}) so much in the cases of nuclei with and without $\Lambda$ hyperon.

\begin{figure}[!hbt]
\includegraphics[bb=0 60 310 220, width=0.6\textwidth]{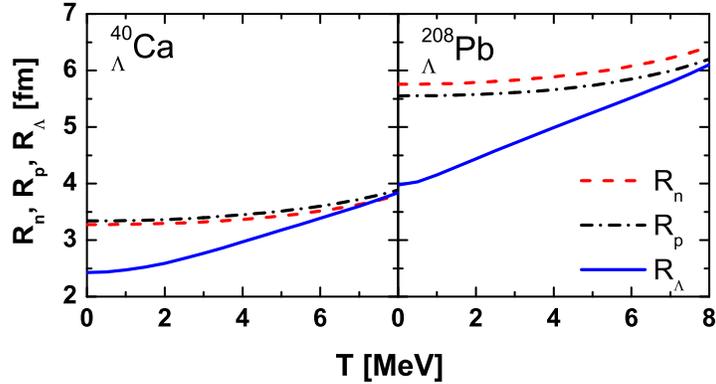}
\caption{(Color online) The rms radii of neutrons, protons, and $\Lambda$ hyperon as a function of temperature $T$ for $^{40}_{\Lambda}$Ca and $^{208}_{\Lambda}$Pb.}\label{ra}
\end{figure}

In Fig.~\ref{ra}, we display the root-mean-square (rms) radii of neutrons, protons and $\Lambda$ hyperon, $R_n,~R_p$ and $R_\Lambda$, as a function of the temperature $T$ for $^{40}_\Lambda$Ca (left panel) and $^{208}_\Lambda$Pb (right panel), which are defined as,
\beq
R_i=\sqrt{\frac{\int d^3rr^2n_i(r)}{\int d^3rn_i(r)}},~~~i=n,~p,~\Lambda.
\eeq 
It is shown that $R_n$ and $R_p$ slowly increase with temperature due to the diffusion of nuclear densities at high temperature. However, the radii of $\Lambda$ hyperon at low temperature are much smaller than the ones of neutrons and protons, while they are very close at high temperature. This is because the $\Lambda$ density distribution becomes much diffuser at high temperature and is more easily influenced by temperature as discussed above.

\begin{figure}[!hbt]
\includegraphics[bb=0 15 330 270, width=0.5\textwidth]{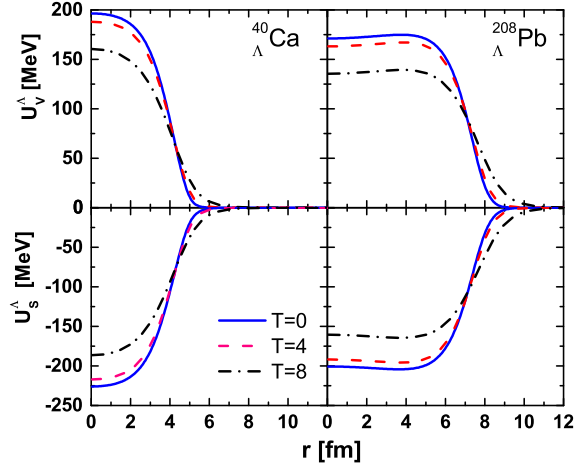}
\caption{(Color online) The scalar and vector $\Lambda N$ potentials as a function of hypernuclei radius at $T=0,~4$ ,and $8$ MeV. The results of $^{40}_{\Lambda}$Ca and $^{208}_{\Lambda}$Pb are shown in left and right panels, respectively. }\label{pot}
\end{figure}

The scalar and vector potentials of $\Lambda$ hyperon in $^{40}_{\Lambda}$Ca and $^{208}_{\Lambda}$Pb at  $T=0,~4,$ and $8$ MeV are shown in Fig. \ref{pot}, which are defined as $U^\Lambda_S=g_{\sigma\Lambda}\sigma$ and $U^\Lambda_V=g_{\omega\Lambda}\omega$. The magnitudes of the scalar and vector potentials reduce with temperature significantly. Especially, at higher temperature, this tendency becomes more obvious. These potentials in the center regions of hypernuclei at $T=8$ MeV are reduced by $20\%$ compared to the cases at $T=0$ MeV. The attractive scalar potential is slightly larger than the repulsive vector potential, and their differences at the center of hypernuclei are about $15-25$ MeV, which result in the bound states of $\Lambda$ hypernuclei.

\begin{figure}[!hbt]
\includegraphics[bb=0 60 310 220, width=0.6\textwidth]{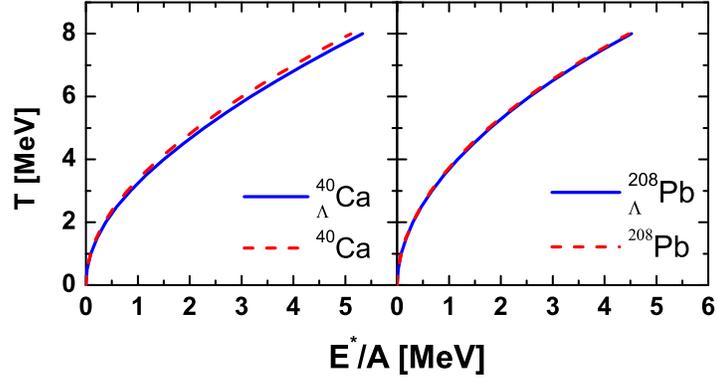}
\caption{(Color online) The caloric curves, i.e., the temperature $T$ as functions of the excitation energy per particle $E^{*}/A$ for $^{40}$Ca, $^{40}_{\Lambda}$Ca, $^{208}$Pb, and $^{208}_{\Lambda}$Pb.}\label{cal}
\end{figure}

The excitation energies of hot $\Lambda$ hypernuclei can be calculated from Eq.~(\ref{eq:Estar}). The temperature $T$ as functions of the excitation energy per particle $E^{\ast}/A$ (caloric curve) are plotted in Fig.~\ref{cal} for $^{40}$Ca, $^{40}_{\Lambda}$Ca, $^{208}$Pb, and $^{208}_{\Lambda}$Pb. We can see that $E^{\ast}/A$ increases slowly at low temperature, while it rises more rapidly as $T$ increases. The excitation energy of $\Lambda$ hypernuclei is larger than the one of normal nuclei with the same baryon number. This is mainly because the single-$\Lambda$ hyperon is more easily excited than a nucleon which has more correlation with other nucleons considering a nucleus as a collective mode. By the same reason, the excitation energy of heavy nuclei is smaller than that of light nuclei at same temperature. We also find that there exists a limiting temperature $T_{\rm{lim}}$ for a hot hypernucleus, which is strongly dependent on the interaction and the size of the box. Generally, the limiting temperature is above $8$ MeV in the Thomas-Fermi calculations \cite{TF87, zhang14b}. Therefore, the results of hot $\Lambda$ hypernuclei in the present work are only shown up to $T \sim 8$ MeV.

\begin{figure}[!hbt]
\includegraphics[bb=60 60 360 220, width=0.6\textwidth]{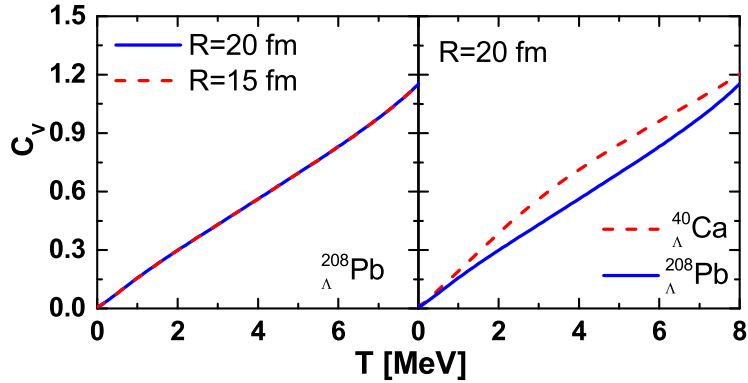}
\caption{(Color online) The specific heat $C_v$ as a function of temperature for $^{208}_{\Lambda}$Pb calculated with different box sizes $R=15$ and $20$ fm (left panel) and those for $^{40}_{\Lambda}$Ca and $^{208}_{\Lambda}$Pb with $R=20$ fm (right panel). }\label{cv}
\end{figure}

The specific heat $C_v$ per particle is a very useful thermodynamic quantity for hot nucleus, which is defined at a fixed volume as,
\beq
C_v=\left.\frac{d(E^{\ast}(T)/A)}{dT}\right|_V.
\eeq
We show in Fig.~\ref{cv} the specific heat as functions of temperature for $^{40}_{\Lambda}$Ca and $^{208}_{\Lambda}$Pb. In Ref. \cite{RTF01}, the specific heat was studied with relativistic Thomas-Fermi approximation for hot nuclei, by introducing a freeze-out volume to treat the density diffusing in the surface of nuclei at finite temperature. Therefore, the specific heat was strongly dependent on the freeze-out volume. When the subtraction procedure is used to isolate the $\Lambda$ hypernucleus from the surrounding baryon gas, the properties of hot hypernucleus are independent of the size of the box. In the left panel of Fig. \ref{cv}, the specific heat of $^{208}_{\Lambda}$Pb is shown with different box sizes $R=15$ fm and $R=20$ fm. We can find that they are identical until $T=8$ MeV. The results of specific heat for $^{40}_{\Lambda}$Ca and $^{208}_{\Lambda}$Pb are compared in the right panel. It is shown that $C_v$ of $^{40}_{\Lambda}$Ca is larger than the one of $^{208}_{\Lambda}$Pb. This is because the caloric curve of $^{40}_{\Lambda}$Ca is stiffer. It is demonstrated that light hypernuclei are more easily excited than heavy one.

The properties of single-$\Lambda$ hypernuclei, $\Lambda$ hyperon radii, center density of $\Lambda$ hyperon, excitation energy per particle, single-$\Lambda$ binding energy, and the level density parameter for $^{40}_{\Lambda}$Ca and $^{208}_{\Lambda}$Pb at different temperatures are listed in Table \ref{pca} and Table \ref{ppb}, respectively. In the low-temperature Fermi gas approximation, the level density parameters $a$, which is related to the density of state of an excited state,  can be expressed as, $S/2T, ~E^{*}/T^2$ , or $S^2/4E^{*}$ \cite{BLV85}, where $E^*$ is the excitation energies from Eq. (\ref{eq:Estar}). In our calculation, the level density parameters with different definitions in single-$\Lambda$ hypernuclei are almost temperature independent for $T\leq 4$ MeV. Their magnitudes are also consistent with each other. The level density parameter for the light nucleus is smaller than the heavy one.

\begin{table}[!htb]
\begin{center}
\resizebox{\textwidth}{!}{
\begin{tabular}{c c c c c c c c }
\hline\hline
             {$T$ (MeV)}&~  {$R_\Lambda$ (fm)}   &~ {$\rho_\Lambda(0)$ ($10^{-2}$fm$^{-3}$)}   &~{$E^{*}/A$ (MeV)}   &~{$B_\Lambda$ (MeV)}  &~ {$S/2T$}  &~ {$E^{*}/T^2$}  &~{$S^2/4E^{*}$}\\
\hline
             {0.0}      &  {$2.43$}               &   {$1.06 $}          &  {$0.00$}             & {$22.19$}            & {$  -   $}  & {$  -   $}       & {$      -    $} \\
             {2.0}      &  {$2.59$}               &   {$0.95 $}          &  {$0.38$}             & {$20.72$}            & {$3.82$}  & {$3.83$}       & {$3.81$} \\
             {4.0}      &  {$2.97$}               &   {$0.73 $}          &  {$1.49$}             & {$17.12$}            & {$3.78$}  & {$3.75$}       & {$3.81$} \\
             {6.0}      &  {$3.38$}               &   {$0.55 $}          &  {$3.18$}             & {$13.01$}            & {$3.64$}  & {$3.53$}       & {$3.75$} \\
             {8.0}      &  {$3.84$}               &   {$0.39 $}          &  {$5.33$}             & {$9.15$}             & {$3.51$}  & {$3.33$}       & {$3.68$} \\
 \hline\hline
\end{tabular}}
\caption{The properties of $^{40}_{\Lambda}$Ca at different temperatures.}\label{pca}
\end{center}
\end{table}

\begin{table}[!htb]
\begin{center}
\begin{tabular}{c c c c c c c c }
\hline\hline
             {$T$ (MeV)}&~  {$R_\Lambda$ (fm)}   &~ {$\rho_\Lambda(0)$ ($10^{-2}$fm$^{-3}$)}   &~{$E^{*}/A$ (MeV)}   &~{$B_\Lambda$ (MeV)}  &~ {$S/2T$}  &~ {$E^{*}/T^2$}  &~{$S^2/4E^{*}$}\\
\hline
             {0.0}      &  {$3.98$}               &   {$0.20 $}        &  {$0.00$}             & {$27.41$}            & {$  -   $}  & {$   -  $}       & {$     -     $} \\
             {2.0}      &  {$4.44$}               &   {$0.15 $}        &  {$0.31$}             & {$25.03$}            & {$16.14$}  & {$15.97$}       & {$16.30$} \\
             {4.0}      &  {$4.99$}               &   {$0.11 $}        &  {$1.17$}             & {$21.13$}            & {$15.55$}  & {$15.19$}       & {$15.93$} \\
             {6.0}      &  {$5.52$}               &   {$0.084 $}        &  {$2.56$}             & {$17.03$}            & {$15.19$}  & {$14.78$}       & {$15.61$} \\
             {8.0}      &  {$6.10$}               &   {$0.064 $}        &  {$4.52$}             & {$13.27$}            & {$15.03$}  & {$14.69$}       & {$15.39$} \\
 \hline\hline
\end{tabular}
\caption{The properties of $^{208}_{\Lambda}$Pb at different temperatures.}\label{ppb}
\end{center}
\end{table}

\begin{figure}[!hbt]
\includegraphics[bb=0 10 320 230, width=0.5\textwidth]{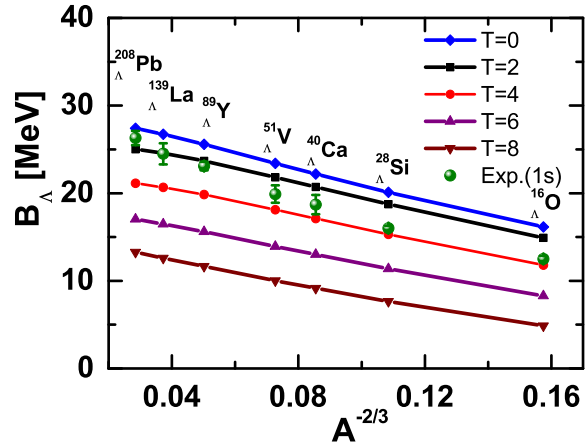}
\caption{(Color online) The single-$\Lambda$ binding energies from $^{16}_{\Lambda}$O to $^{208}_{\Lambda}$Pb at different temperatures $T=0,~2,~4,~6$, and $8$ MeV and compared with the experimental data for $1s$ states at zero temperature~\cite{hashimoto06}.}\label{blt}
\end{figure}

To distinguish the excitation energy in Eq. (\ref{eq:Estar}), we would like to use the $\Lambda$ binding energy instead of $\Lambda$ excitation energy in this paper, although the Lambda hyperon may also occupy an excited state. The single-$\Lambda$ binding energy is a very important property of $\Lambda$ hypernuclei, which is obtained by the subtraction of the binding energy of $\Lambda$ hypernucleus from its core energy without hyperon. In Fig. \ref{blt}, we present the single-$\Lambda$ binding energies of some typical spherical single-$\Lambda$ hypernuclei at different temperatures from $^{16}_{\Lambda}$O to $^{208}_{\Lambda}$Pb and compare them with the experimental data in term of $\Lambda$ $1s$ states at zero temperature~\cite{hashimoto06}. It is seen that the single-$\Lambda $ binding energies decrease with temperature. At high temperature, such reduction becomes faster. At $T=0$ MeV, the single-$\Lambda$ binding energy can be measured at different spin states. The experimental value of the $1s$ state of $^{208}_{\Lambda}$Pb is $26.3\pm0.8$ MeV \cite{hashimoto06}. In the present study, we use the relativistic Thomas-Fermi approximation to describe the hypernucleus and we do not solve the Dirac equation for the nucleon and $\Lambda$ hyperon. Therefore, the single-$\Lambda$ binding energies in this approximation cannot be distinguished from different spin states. The $\Lambda$ binding energy of $^{208}_{\Lambda}$Pb obtained in the present calculation at $T=0$ MeV is 27.41 MeV, which is consistent with the experiment data. For the light hypernuclei, like $^{16}_{\Lambda}$O, our results of $\Lambda$ binding energies are slightly overestimated in comparison with experimental data.

\section{Conclusion}
The relativistic Thomas-Fermi approximation has been applied to the investigation of hot single-$\Lambda$ hypernuclei using the RMF model for the interaction of baryons. The subtraction procedure has been employed in order to separate the hypernucleus from the surrounding baryon gas. With such treatment, the properties of hot $\Lambda$ hypernucleus are independent of the size of the box in which the calculation is performed. The nucleon and $\Lambda$ hyperon interact via the exchange of the $\sigma$ and $\omega$ mesons, whose coupling constants are determined by experimental $\Lambda$ binding energies in the RMF model.

 We have studied two single-$\Lambda$ hypernuclei, $^{40}_{\Lambda}$Ca and $^{208}_{\Lambda}$Pb, as numerical examples in this work. At high density, the $\Lambda$ gas density becomes visible and increases with temperature. On the other hand, the $\Lambda$ density at the center of $\Lambda$ hypernuclei is reduced largely with temperature. The temperature dependence of $\Lambda$ densities is more remarkable than that of proton and neutron, since one hyperon is more easily excited than nucleus which are compounded of many nucleons. Furthermore, the magnitudes of $\Lambda$ densities at the center of hypernuclei are almost inverse to the baryon numbers. The rms radius of $\Lambda$ hyperon is clearly different from those of proton and neutron at zero temperature. However, it increases rapidly with temperature and becomes comparable with the radii of proton and neutron, which is due to the diffusion of $\Lambda$ distribution at high temperature. The scalar and vector potentials of $\Lambda$ hyperon have been found to be reduced with temperature so that the $\Lambda$ binding energies become small at high temperature. The specific heat defined as the derivation of excitation energy with respect to temperature was found to be independent of the size of the box by employing the subtraction procedure, which is different from introducing the freeze-out volume to consider the temperature effect. Finally we also gave the single-$\Lambda$ binding energies from $^{16}_{\Lambda}$O to $^{208}_{\Lambda}$Pb at different temperatures. The binding energies are consistent with the results obtained in the RMF model at zero temperature for heavy hypernuclei. As temperature increases, the $\Lambda$ binding energies decrease significantly.

We have systematically studied the properties of hot single-$\Lambda$ hypernuclei above mediate mass. There are also some experimental data for light single-$\Lambda$ hypernuclei, double-$\Lambda$ hypernuclei, and $\Xi$ hypernuclei. Further work is required to investigate the properties of various hypernuclei at finite temperature.

\section*{Acknowledgments}
This work was supported in part by the National Natural Science Foundation of China (Grant No. 11375089 and Grant No. 11405090).

\end{document}